\documentclass[twocolumn,grl]{agutex}

\usepackage{graphicx}
\setkeys{Gin}{draft=false}
\authorrunninghead{DAUB ET AL.}

\titlerunninghead{ARE MEGAQUAKES CLUSTERED?}

\authoraddr{E. Ben-Naim, Theoretical Division, Los Alamos
National Laboratory, MS B213, Los Alamos, NM 87545, USA.
(ebn@lanl.gov)}

\authoraddr{E. G. Daub, R. A. Guyer, P. A. Johnson,
Geophysics Group, Los Alamos National Laboratory, MS D443,
Los Alamos, NM 87545, USA. (edaub@lanl.gov)}

\begin{document}

%
%

\title{Are megaquakes clustered?}
%
%

%
%



\authors{Eric G. Daub,\altaffilmark{1,2}
Eli Ben-Naim,\altaffilmark{2,3} Robert A. Guyer,\altaffilmark{1,4} 
and Paul A. Johnson\altaffilmark{1}}

\altaffiltext{1}{Earth and Environmental Sciences Division, Los Alamos National Laboratory,
Los Alamos, New Mexico, USA.}

\altaffiltext{2}{Center for Nonlinear Studies, Los Alamos National Laboratory,
Los Alamos, New Mexico, USA.}

\altaffiltext{3}{Theoretical Division, Los Alamos National Laboratory,
Los Alamos, New Mexico, USA.}

\altaffiltext{4}{Physics Department, University of Nevada, Reno, Nevada, USA.}

%
%


\begin{abstract}
We study statistical properties of the number of large earthquakes
over the past century. We analyze the cumulative
distribution of the number of earthquakes with magnitude larger than
threshold $M$ in time interval $T$, and quantify the statistical significance
of these results by simulating a large number of synthetic random
catalogs. We find that in general, the earthquake record cannot be distinguished from a process
that is random in time. This
conclusion holds whether aftershocks are removed or not, except at magnitudes
below $M=7.3$.
At long time intervals ($T$ = 2-5~years), we find that
statistically significant clustering is present in the catalog for lower magnitude
thresholds ($M$ = 7-7.2). However, this clustering
is due to a large number of earthquakes on record in the early part of the 20th century,
when magnitudes are less certain.
\end{abstract}

%
%

%

\begin{article}

%
%

\section{Introduction}
\begin{figure*}
\noindent\includegraphics[width=39pc]{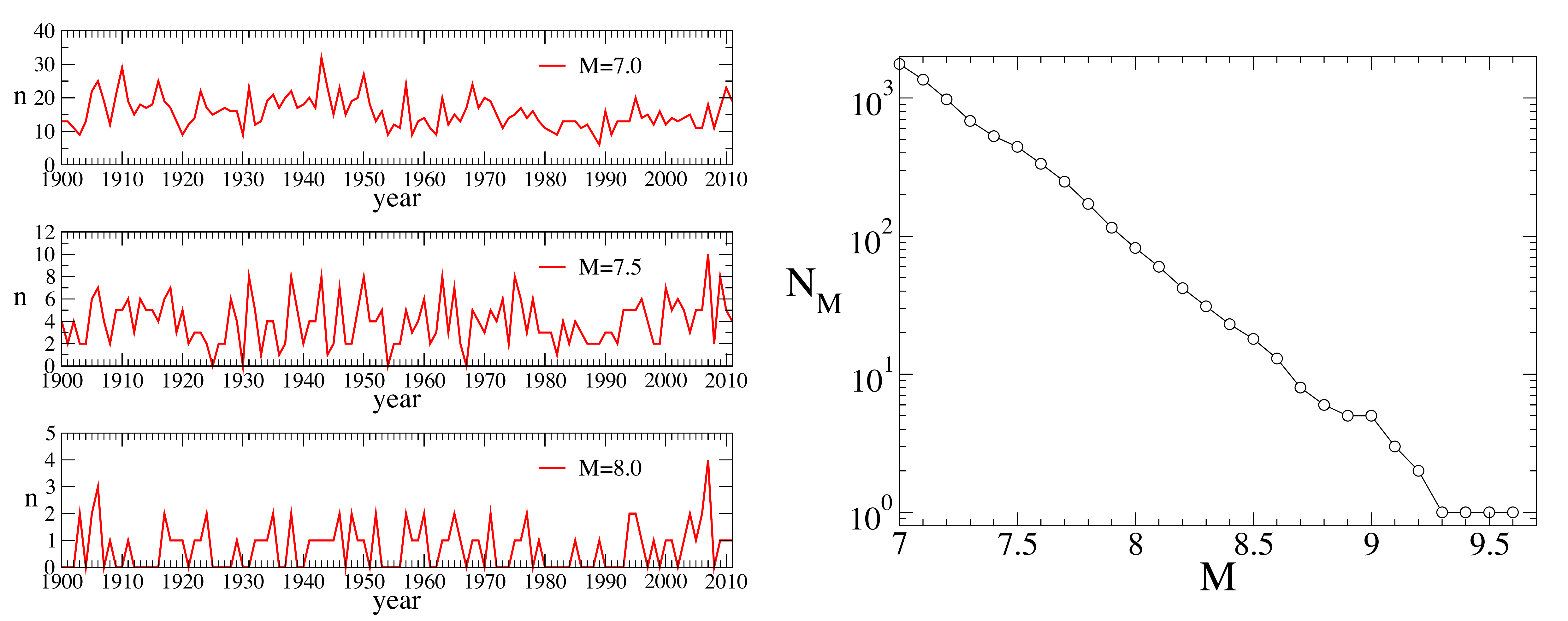}
\caption{(left) The number of large earthquakes, $n$, in a calendar year over
the past century (1900-2011). Here, large earthquakes are defined as
events with magnitude greater than or equal to $M$. Three thresholds
were used: $M=7.0$ (top), $M=7.5$ (middle), and $M=8.0$ (bottom).
(right) The cumulative number $N_M$ of large earthquakes with
magnitude of at least $M$ during the time period $1900-2011$.
\label{fig-nt}}
\end{figure*}

The number of powerful earthquakes worldwide has increased over
the past decade (Fig.~\ref{fig-nt} (left)). This increase has prompted debate whether
large earthquakes cluster in time [{\it Kerr}, 2011]. If so,
this would have an impact on how seismic hazard is assessed
worldwide. Multiple studies have investigated this
question [{\it Bufe and Perkins}, 2005; {\it Brodsky}, 2009; {\it Michael}, 2011; {\it Shearer and Stark}, 2012; {\it Ammon et al.}, 2011; {\it Bufe and Perkins}, 2011]. Conclusions have been mixed, with some studies finding evidence
of clustering [{\it Bufe and Perkins}, 2005; 2011], while others have concluded that earthquakes
cannot be distinguished from a process that is
random in time [{\it Michael}, 2011; {\it Shearer and Stark}, 2012].

In parallel, recent studies show that earthquakes can be dynamically triggered by
seismic waves [{\it Hill et al.}, 1993; {\it Gomberg et al.}, 2004; {\it Freed}, 2005]. 
It is not clear if large earthquakes can trigger other large
earthquakes; one recent study
did not find evidence of such triggering [{\it Parsons and Velasco}, 2011],
although this remains an open question in seismology. If large earthquakes do cluster in time, this might suggest
that large earthquakes can be dynamically triggered.

We study the statistics of large ($M\geq7$) earthquakes from 1900-2011
to assess whether earthquakes deviate from random occurrence. We examine
the catalog both with and without removal of aftershocks, and use transparent statistical measures
to quantify the likelihood that a random process could produce the earthquake record.

\section{Data and Aftershock Removal}

Our statistical analysis uses the USGS PAGER catalog of large
earthquakes [{\it Allen et al.}, 2009], supplemented with the Global CMT catalog through the
end of 2011.
The catalog consists of 1761 events with
magnitude $M>7.0$.  As can be seen from the magnitude-frequency plot in
Fig.~\ref{fig-nt} (right), this catalog adheres to the ubiquitous
Gutenberg-Richter law [{\it Gutenberg and Richter}, 1954],
and is complete for magnitude $M>7.0$. The magnitudes in the
PAGER catalog are a mix of magnitude types -- the majority of events
are given in moment magnitude, but events early in the century often use
a different magnitude measure, such as surface wave magnitude.
Because very large earthquakes are rare, any study of the statistics
of this dataset is inherently limited by the small number of extremely
powerful earthquakes on record.

We have studied two additional catalogs, one compiled by {\it Pacheco and Sykes} [1992],
and one based on the NOAA Significant Earthquake Database
(National Geophysical Data Center/World Data Center (NGDC/WDC) Significant Earthquake Database, Boulder, CO, USA, available at 
http://www.ngdc.noaa.gov/hazard/earthqk.shtml). We find that the results depend 
on the catalog choice due to discrepancies in magnitude between the catalogs. Because
PAGER contains more events, and the magnitudes in PAGER are the most consistent
with the Gutenberg-Richter Law, we focus on PAGER in our analysis. A comprehensive
study of the discrepancies between catalogs will be the subject of future work.

While the PAGER catalog is the most complete record of large earthquakes,
the data has limitations. First, because seismic instruments were relatively sparse
in the first half of the 20th century, data for these events have larger uncertainties.
Additionally, the data includes aftershocks.
Aftershock removal is not trivial, and it requires assumptions that
cannot be tested rigorously due to limited data.

We remove aftershocks by flagging any event within
a specified time and distance window of a larger magnitude main shock [{\it Gardner and Knopoff}, 1974].
We use the time window from the original {\it Gardner and Knopoff}
study. The distance window should be similar to the rupture length of the main shock. However,
rupture length data does not exist for the entire catalog.
Therefore, we must estimate the rupture length based on magnitude. This is problematic
because the catalog contains multiple
types of faulting (i.e. subduction megathrust, crustal strike-slip, etc.), each
with a different typical rupture length for a given magnitude.
For example,
the 2002 $M=7.9$ Denali earthquake and the 2011 $M=9.0$ Tohoku Earthquake did not have
substantially different rupture lengths
[{\it Eberhart-Philips et al.}, 2003; {\it Simons et al.}, 2011] despite a large difference in 
seismic moment. We use an empirical rupture length formula
[{\it Wells and Coppersmith}, 1994], and
choose to be conservative by doubling the {\it Wells and Coppersmith} subsurface rupture length 
estimate for reverse faulting. We have studied various choices for this rupture length multiplicative
factor, and find that doubling the rupture length estimate makes the rupture lengths large enough
to be fairly conservative, but not so large as to excessively remove events from the catalog.
This may remove some events from the catalog that are not aftershocks,
but it will not bias our results by leaving many aftershocks in the catalog.
After removal of aftershocks,
the PAGER catalog is reduced to 1253 events. In this investigation, we first
examine the entire catalog to draw as much information from the raw
data as possible before introducing assumptions about aftershocks.

\section{Statistical Analysis}

Our study utilizes the cumulative probability
distribution of the number of large earthquakes in a fixed time
interval $Q_n$. The cumulative distribution gives the probability that there
are at least $n$ earthquakes with magnitude of at least $M$ in a given time
interval $T$, measured in months.
We compare the 
observed frequency distribution
$Q_n$ with the frequency distribution for a random Poisson process. 
Let the
average number of large earthquakes in a time interval be $\alpha$. If large earthquakes are not correlated in
time, then the probability $P^{{\rm rand}}_{n}$ that there are $n$
events during a time interval is
\begin{equation}
P^{{\rm rand}}_{n}=\frac{\alpha^n}{n!}e^{-\alpha}.
\end{equation}
The Poisson distribution is characterized by a single parameter, the
average. We also note that the average and the variance are identical,
$\langle n\rangle =\langle n^2\rangle -\langle n\rangle^2=\alpha$.
The cumulative distribution for a Poissonian catalog $Q^{{\rm
rand}}_n$ is given by the following sum:
\begin{equation}
\label{qn}
Q^{{\rm rand}}_{n}=\sum_{m=n}^\infty P^{\rm rand}_m=\sum_{m=n}^\infty\frac{\alpha^m}{m!}e^{-\alpha}.
\end{equation}
Note that $Q^{\rm rand}_n$ depends on the choice of $M$ and $T$, as
these determine the average event rate $\alpha$.  We calculate $Q_n$
for the earthquake data, and compare the data with the expected
distribution for a Poissonian catalog $Q^{{\rm rand}}_n$. Note that the cumulative distribution
forms the basis of one of the statistical tests used in {\it Shearer and Stark} [2012], but here
we explore many time bin sizes to see if the results depend on the choice of the time window.

\begin{figure*}
\includegraphics[width=39pc]{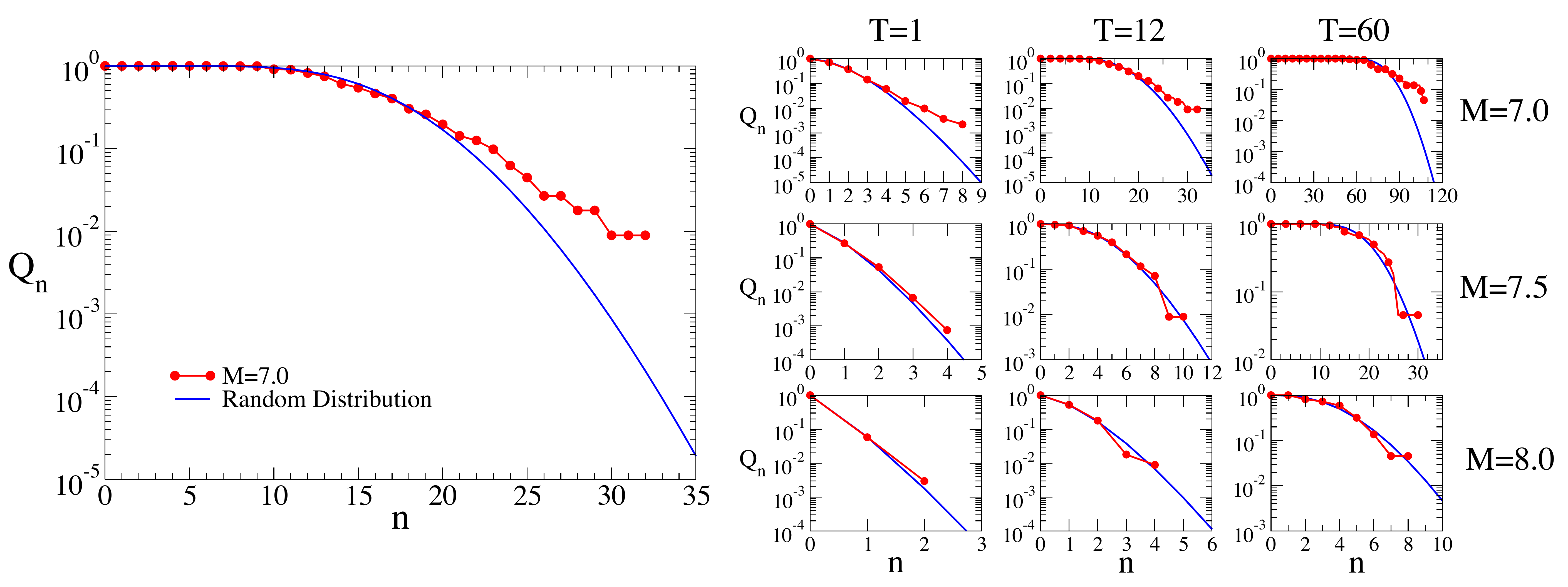}
\caption{The cumulative frequency distribution at different threshold magnitudes
and time intervals. (left) $Q_n$ versus $n$ for $M=7.0$ and $T=12$~months, compared to
the distribution expected for a random catalog. (right) $Q_n$ versus $n$, obtained using
magnitude thresholds $M=7.0$ (top), $M=7.5$ (middle), and $M=8.0$
(bottom) and time intervals $T=1$~month (left), $T=12$~months (middle), and $T=60$~months
(right). The solid lines indicate the expected distribution for a Poissonian catalog.}
\label{fig-panel9}
\end{figure*}

Figure \ref{fig-panel9} (left) shows an example of the cumulative distribution plot for the raw PAGER catalog
for $M=7$ and $T=12$~months. The cumulative distribution $Q_n$ quantifies the probability that a time
window contains at least $n$ events. Thus, the curves always begin at $Q_0=1$, and decrease
as $n$ increases. The final point on each plot corresponds to
the maximum number of events observed in the 
chosen time window.

Figure \ref{fig-panel9} (left) shows that the frequency of large
earthquakes with $M\geq 7.0$ is roughly Poissonian below the
average $\alpha=15.7$ events/year. However, the tail of the cumulative distribution
is {\em overpopulated} with respect to the Poisson
distribution. An overpopulated tail indicates that events are clustered in time. We perform this analysis for higher
magnitude thresholds ($M=7.5$, $M=8$) and both longer and shorter time window sizes
($T=1$~month, $T=60$~months), and the results are shown in Fig.~\ref{fig-panel9} (right).
The bins evenly divide the catalog into
an integer number of fixed time windows: $T=1$~month corresponds to $112\times12=1344$ bins, 
and $T=12$~months corresponds to 112 bins. For $T=60$~months, the catalog cannot be evenly
divided into 5 year bins. Therefore, it is instead 
divided into the closest integer number of bins (22), which means that the bin size is actually slightly larger than 60 months.

We find that the catalog exhibits an overpopulated tail only for $M=7$.
Within the $M=7$ data, the overpopulation
is found for all $T$. The strength of this overpopulation is significant because it
can be a few orders of magnitude. However, the catalog at
$M=7.5$ and $M=8$ agrees very well with the prediction for a Poissonian catalog. This is remarkable, as even with a relatively small number
of earthquakes, the data is in agreement with a random distribution.

\begin{figure*}
\includegraphics[width=39pc]{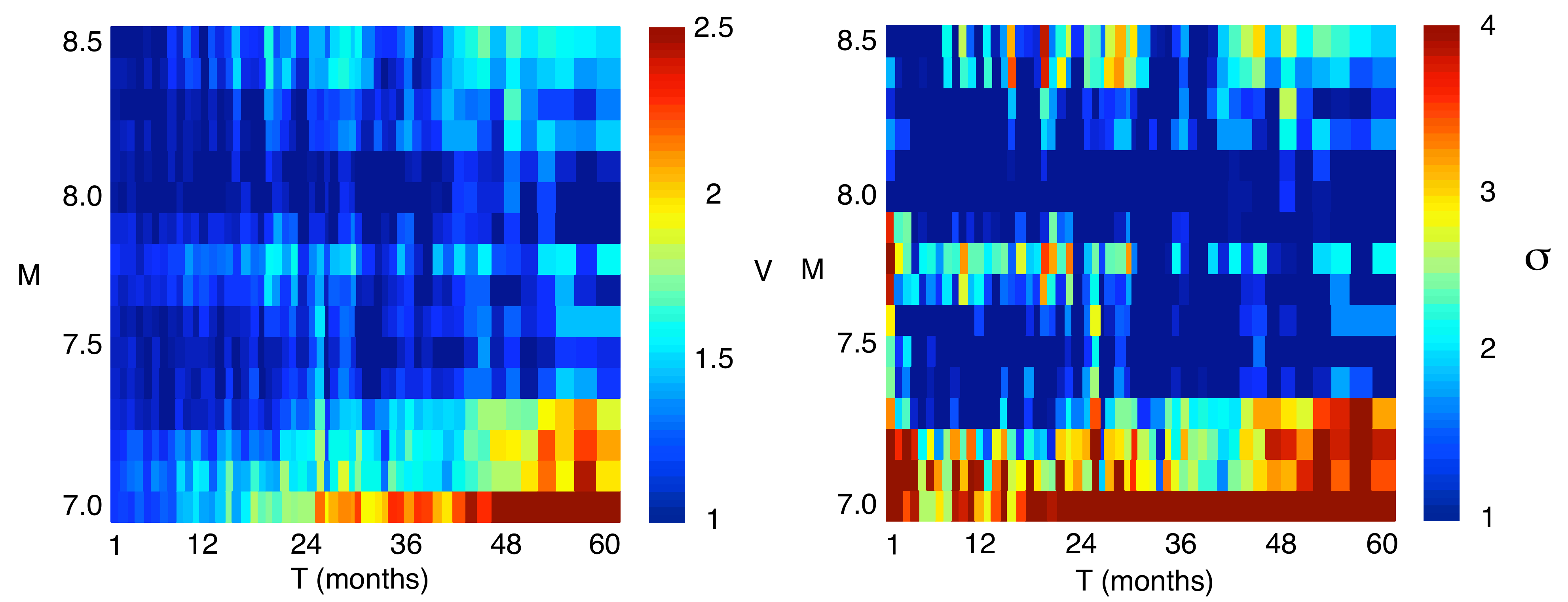}
\caption{(left) The normalized variance $V$ versus magnitude threshold
$M$ and the time interval $T$ (in months). The normalized variance is
color coded with red indicating strong overpopulation and blue
indicating a random distribution. (right) Standard deviations above the mean variance $\sigma$
as a function of $M$ and $T$, determined from $10^6$ Poissonian synthetic catalogs. Again,
statistically significant overpopulation is indicated in red, while blue indicates a random distribution.}
\label{fig-var}
\end{figure*}

To quantify the statistical significance of the overpopulation, we utilize the normalized
variance:
\begin{equation}
V=\frac{\langle n^2\rangle-\langle n\rangle^2}{\langle n\rangle}.
\end{equation}
An observed distribution with a strongly overpopulated tail necessarily has a
large variance. Moreover, a value close to unity is expected for a catalog that is
random in time, while a value larger than unity indicates clustering.
Hence, the normalized variance $V$ is a
convenient, scalar, measure of clustering.
The normalized variance is shown as a function of $M$ and $T$
in Fig.~\ref{fig-var} (left), and confirms that at $M=7$ the catalog is clustered.
In this analysis, we compute $V$ with many different bin sizes, ranging from 1 month
up to approximately 5 years. In each case, the
number of bins is chosen to be an integer so that we always utilize the entire catalog
(i.e. the time bin size is not always an integer number of months).

To test whether the clustering observed in the data is statistically significant,
we generate $10^6$ synthetic Poissonian catalogs
with an average event rate given by $\alpha=1761/112$ events/year, the same as
in the PAGER catalog. Each event is
assigned a magnitude, drawn randomly from the actual catalog magnitudes
with replacement. Using the $10^6$ Poissonian realizations of the earthquake catalog,
we compute the average normalized variance $\bar{V}$ and the standard deviation
of the normalized variance $\delta V$ as a function of $M$ and $T$.
Conveniently, the normalized variance for an ensemble of synthetic random catalogs
is approximately described by a normal distribution. This makes this quantity
useful for determining the statistical significance of the observed clustering.
The normalized variance determined from the earthquake data $V$ 
can then be expressed as a certain number of
standard deviations above the mean $\sigma$,
\begin{equation}
\sigma=\frac{V-\bar{V}}{\delta V}.
\end{equation}
Since $V$ is normally distributed for an ensemble of random catalogs, we know that if
the value of $V$ determined from the data is larger than $\bar{V}$ by
several standard deviations, this indicates that the catalog contains statistically significant
clustering. 

The number of standard deviations above the mean $\sigma$ is shown as a function of $M$
and $T$ in Fig.~\ref{fig-var} (right). In the plot,
red indicates statistically significant clustering, and blue indicates a variance consistent with a
random catalog.
This analysis verifies the results from the cumulative distribution: clustering is observed
at low magnitudes ($M<7.3$), while no significant clustering is observed at higher magnitudes
($M\geq7.3$). This observation is robust over time bin sizes ranging from 1 month
to 5 years.
Note that while the normalized variance is much larger for $M=7$ and $T=60$~months
than for $M=7$ and $T=1$~month, in both cases the normalized variance is several standard
deviations above the mean. This is because there is more variability in the normalized
variance for longer time bins -- we find
that $\delta V\sim T^{1/2}$, independent of the magnitude threshold.
We stress that our analysis thus far relies
on the complete earthquake record which necessarily includes
aftershocks. Hence, aftershock removal is not even necessary to
demonstrate that the statistics of large earthquakes with magnitude
$M>7.3$ show no significant clustering.

\begin{figure}
\includegraphics[width=19pc]{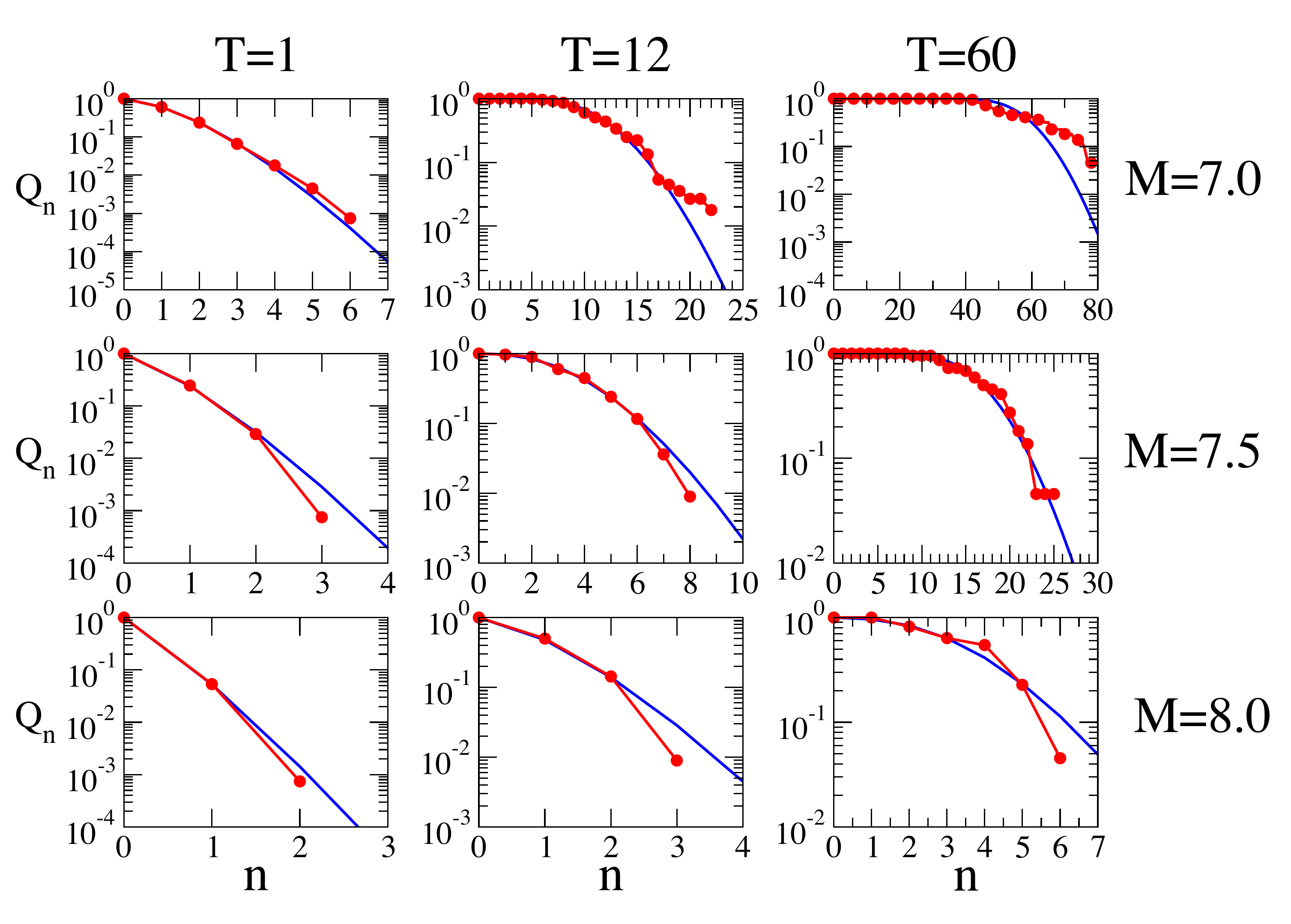}
\caption{The cumulative frequency distribution for the catalog with aftershocks removed at different threshold
magnitudes and time intervals. Shown is $Q_n$ versus $n$, obtained using
magnitude thresholds $M=7.0$ (top), $M=7.5$ (middle), and $M=8.0$
(bottom) and time intervals $T=1$~month (left), $T=12$~months (middle), and $T=60$~months
(right). The solid lines indicate the expected distribution for a Poissonian catalog.}
\label{fig-panel9declust}
\end{figure}

We repeat the above analysis, with aftershocks
removed, to test if the clustering observed for $M<7.3$ is due to
aftershocks.
The results of the cumulative distribution analysis with aftershocks removed
is shown in Fig.~\ref{fig-panel9declust}.
The catalog now closely follows the cumulative distribution for a Poissonian catalog
for $M=7$, $T=1$~month, demonstrating
that the clustering at short times and lower magnitudes is due to aftershocks. There is still
overpopulation for $M=7$ at longer times. At higher magnitudes, many of the
curves appear slightly underpopulated for large numbers of events. This could be due to our
conservative aftershock removal procedure, which may have removed some independent
events.

\begin{figure*}
\includegraphics[width=39pc]{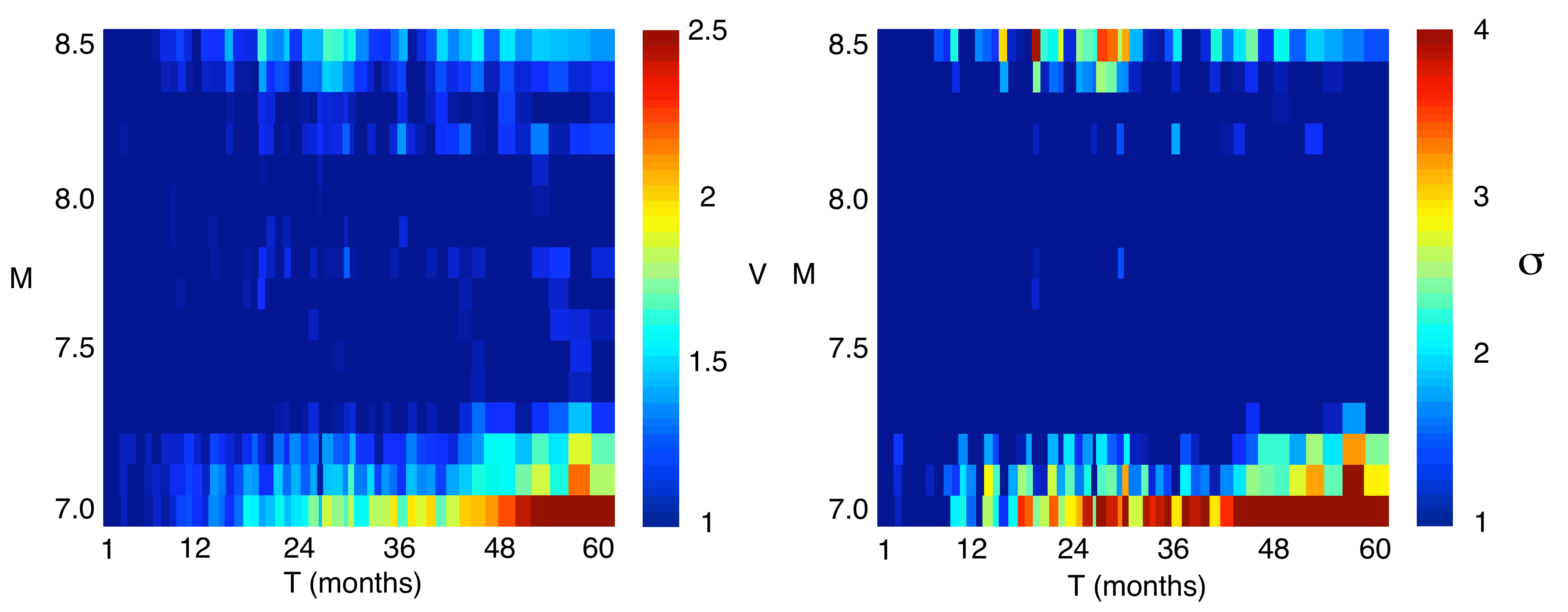}
\caption{(left) The normalized variance $V$ versus magnitude threshold
$M$ and the time interval $T$ (in months) for the catalog with aftershocks removed. The normalized variance is
color coded with red indicating strong overpopulation and blue
indicating a random distribution. (right) Standard deviations above the mean variance $\sigma$
for the catalog with aftershocks removed as a function of $M$ and $T$, determined from $10^6$ Poissonian synthetic catalogs. Again,
statistically significant overpopulation is indicated in red, while blue indicates a random distribution.}
\label{fig-vardeclust}
\end{figure*}

Calculations using synthetic catalogs and the variance measure $V$ confirm these results. 
Figure~\ref{fig-vardeclust} shows that the clustering observed for small magnitudes
($M<7.3$) and short times
($T<12$~months) no longer occurs once aftershocks are removed from the catalog.
Interestingly, the clustering at longer time intervals ($T>24$ months) persists.
Most likely, this clustering is due to the
fact that there is a mismatch between the event rates in the first and
the second halves of the century, the former being larger by about
$20\%$.
This can be seen in Fig.~\ref{fig-nt} (left, top), which shows several spikes in the number
of $M\geq7$ events during the first half of the century. If we divide the catalog
into two time periods (1900-1955 and 1956-2011), we find that each half of the data is consistent
with random earthquake occurrence, with a different rate for each half.
Because magnitude estimates early in the century are subject to larger uncertainties and may be
systematically overestimated [{\it Engdahl and Villase\~{n}or}, 2002], 
it is not clear if this clustering is real or due to less reliable data.

\section{Conclusions}

Our studies using the PAGER earthquake catalog demonstrate
that the catalog cannot be distinguished from random earthquake occurrence. This is in
agreement with several other recent studies [{\it Michael}, 2011; {\it Shearer and Stark}, 2012]. We do find evidence
of clustering for $M=7$ and $T=2$-5 years, which was not identified in
the other studies. However, we note that this clustering is due to a large
number of events on record early in the 20th Century.

For large events ($M>7.3$), the catalog with aftershocks is well described by a process that is
random in time. This is because large aftershocks are rare, and there are relatively few
large events in the catalog to begin with.
Because clustering due to aftershocks, which is known to be present in the data,
is not detectable by our
statistical tests, it is possible that there is clustering in the catalog at large magnitudes that is obscured
by the small amount of data. Future studies will examine the likelihood of identifying clustering
in synthetic clustered catalogs given the small amount of data in the earthquake catalog.

These findings underscore that we have very little megaquake data, due to limited instrumentation.
Increases in the number of seismic and geodetic instruments in recent years
has led not only to the improved identification and characterization of
large earthquakes, but also to the discovery of novel slip behaviors such as low frequency earthquakes [{\it Katsumata and Kamaya} 2003],
very low frequency earthquakes [{\it Ito et al.}, 2006], slow slip events [{\it Dragert et al.}, 2001], and
silent earthquakes [{\it Kawasaki et al.}, 1995].
Integrating observations of other types of events with earthquake
data may prove to be the key to identifying causal links between events,
providing a comprehensive picture of the interactions that may underlie the
physics of great earthquakes.


%
%
%
%
%
%
%

\begin{acknowledgments}
The USGS PAGER catalog is available on the web at http://earthquake.usgs.gov/earthquakes/pager/, and the Global CMT catalog is available at http://www.globalcmt.org/. We thank Terry Wallace, Thorne Lay, Charles Ammon, and Joan Gomberg for useful comments.  
This research has been supported by DOE grant DE-AC52-06NA25396 and institutional
(LDRD) funding at Los Alamos.
\end{acknowledgments}

\end{article}


%
%

%
%
%
%
%
%
%


\end{document}